\documentclass[twoside]{article}
\usepackage{multicol,graphicx,amssymb,mathrsfs,amsmath,pifont,amscd,latexsym,amsthm,geometry}
\usepackage{multicol}
\input vatola.sty \input cyracc.def
\input psfig.sty

\newcommand{\sanhao}{\fontsize{19.08pt}{3\baselineskip}\selectfont}

\newcommand{\wuhao}{\fontsize{10.5pt}{\baselineskip}\selectfont}
\newcommand{\xiaowuhao}{\fontsize{9.5pt}{\baselineskip}\selectfont}

\newcommand{\liuhao}{\fontsize{7.875pt}{\baselineskip}\selectfont}

\newcommand{\qihao}{\fontsize{7pt}{\baselineskip}\selectfont}
\newcommand{\bahao}{\fontsize{8pt}{\baselineskip}\selectfont}
\TagsOnRight

\renewcommand{\baselinestretch}{1.06} 
\catcode`@=11 \long\def\@makefntext#1{\noindent #1}
\newskip\tabcentering \tabcentering=1000pt plus 1000pt minus 1000pt
\def\REF#1{\par\hangindent\parindent\indent\llap{#1\enspace}}
\def\MCH#1#2{\setbox0=\hbox{\raise#1\hbox{#2}}\smash{\box0}}

\def\@evenfoot{}\def\@oddfoot{}
\def\@evenfoot{\vbox{\hbox to \textwidth{\bahao\sf\hbox to
0.01cm{\textbf{\thepage}\hfill} \hfill{\emph{AN ChunSheng et al. Sci
China Ser F-Inf Sci }{$|$ G. 2009 $|$ vol. 52 $|$ no. 10 $|$
\textbf{1452-1457}}}\hfill}}}
\def\@oddfoot{\vbox{\hbox to \textwidth{\bahao\sf\hbox to
0.01cm{} \hfill{ \emph{AN ChunSheng et al. Sci China Ser F-Inf Sci
}{$|$ G. 2009 $|$ vol. 52 $|$ no. 10 $|$
\textbf{1452-1457}}}\hfill\hfill\textbf{\thepage}}}}

  \def\tlj{\end{document}}  \newsymbol\wjzhml 203F

\def\tlj{\end{document}}
\def\no{\noindent}

\begin{document}
\abovedisplayskip=5pt plus 1pt minus 2pt 
\belowdisplayskip=5pt plus 1pt minus 2pt 
\columnsep 20pt
\textwidth=176truemm \textheight=229truemm
\renewcommand{\baselinestretch}{0.9}\baselineskip 9pt

\vspace{11.6true mm}
\renewcommand{\baselinestretch}{0.7}\baselineskip 5pt
\noindent{\parbox{16.3cm}{\sanhao{\sf\textbf{Strong decays of $N^{*}(1535)$ in an extended chiral quark model}}}}
\vspace{0.5 true cm}
\renewcommand{\sfdefault}{phv}

\renewcommand{\baselinestretch}{1.6}\baselineskip 19.08pt
\noindent{\sf  AN ChunSheng$^{1,2,3}$\& ZOU BingSong$^{1,2,3\dag}$
\footnotetext{ \baselineskip 6pt \qihao
\vspace{-2.2mm}\\
Received April 28, 2009; accepted May 23, 2009\\
doi: 10.1007/s11433-009-0199-6\\
$^\dag$Corresponding author (email: zoubs@ihep.ac.cn)\\
Supported by the the National Natural Science Foundation of China
under grants Nos. 10875133, 10821063, and by the Chinese
Academy of Sciences under project No.~KJCX3-SYW-N2.\vspace{2mm}\\

\centerline{\bahao\sf \emph{Sci China Ser F-Inf Sci} $|$ G. 2009 $|$
vol. 52 $|$ no. 10 $|$
\textbf{1452-1457}}}}

\vspace{0.2 true cm}\noindent
\parbox{16.3cm}
{\noindent\renewcommand{\baselinestretch}{1.3}\baselineskip 12pt
{\liuhao\sf $^1$ Institute of High Energy Physics, CAS, P.O.Box 918,
Beijing 100049, China;\\
$^2$ Theoretical Physics Center for Science Facilities, CAS, P.O.Box
918, Beijing 100049, China;\\
$^3$ Graduate School, Chinese Academy of Sciences,
Beijing 100049, China\vspace{2mm}}}

\noindent{\xiaowuhao\sf\textbf{\hspace{-1mm}
\parbox{16.3cm}
{\noindent
\renewcommand{\baselinestretch}{1.3}\baselineskip 13pt
The strong decays of the $N^{*}(1535)$ resonance are investigated in
an extended chiral quark model by including the low-lying
$qqqq\bar{q}$ components in addition to the $qqq$ component. The
results show that these five-quark components in $N^{*}(1535)$
contribute significantly to the $N^{*}(1535)\to N\pi$ and
$N^{*}(1535)\to N\eta$ decays. The contributions to the $N\eta$
decay come from both the lowest energy and the next-to-lowest energy
five-quarks components, while the contributions to the $N\pi$ decay
come from only the latter one. Taking these contributions into
account, the description for the strong decays of $N^{*}(1535)$ is
improved, especially, for the puzzling large ratio of the decays to
$N\eta$ and $N\pi$.}}}

\vspace{5.5mm}{\noindent \footnotesize \sf
\parbox{16.3cm}
{\noindent
\renewcommand{\baselinestretch}{1.3}\baselineskip 13pt
$N^{*}(1535)$, strong decays, five-quark components. }} \vspace{6mm}
\baselineskip 15pt

\begin{multicols}{2}

\renewcommand{\baselinestretch}{1.08}
\parindent=10.8pt  
\rm\wuhao  \vspace{-4mm}

正文


Among the low-lying nucleon excitations, the $S_{11}$ state
$N^{*}(1535)$ plays a special role due to its large $N\eta$ decay
rate$^{[1]}$, even though its mass is very close to the threshold of
the decay. And recently it has been shown that the coupling of
$N^{*}(1535)N\phi$ may be significant$^{[2]}$, which is consistent
with the previous indications of the notable $N^{*}(1535)K\Lambda$
coupling$^{[3]}$ deduced from BES data. These suggest that there are
large $s\bar{s}$ components in the $N^{*}(1535)$ resonance. And in a
recent paper$^{[4]}$, the role of the low-lying $qqqq\bar{q}$
components in the electromagnetic transition $\gamma^{*}N\rightarrow
N^{*}(1535)$ has been investigated, the result shows that the
contributions of the $s\bar{s}$ component in $N^{*}(1535)$ are
significant, and with admixture of 5-quark components with a
proportion of 20\% in the nucleon and 25-65\% in the $N^{*}(1535)$
resonance the calculated helicity amplitude $A^{p}_{1/2}$ decreases
at the photon point, $Q^{2}=0$ to the empirical range. Consequently,
it also suggests large $s\bar{s}$ component in $N^{*}(1535)$, which
is in line with the predictions in Refs.$^{[2-3]}$.

Here we investigate the strong decays of the $N^{*}(1535)$ resonance
by extending the chiral quark model$^{[5-6]}$ to include the
$qqqq\bar{q}$ components, which has been applied to the strong
decays of the $\Delta(1232)$ and $N^{*}(1440)$ resonances
successfully$^{[7-8]}$. The wave functions of the $qqq$ components
in the nucleon and $N^{*}(1535)$ are taken to be the conventional
ones$^{[6]}$, while we employ the wave functions for the
$qqqq\bar{q}$ components in the nucleon given in Ref.$^{[9]}$, and
the orbital-flavor-spin configuration of the four-quark subsystem is
taken to be $[31]_{X}[4]_{FS}[22]_{F}[22]_{S}$, which leads to the
lowest energy. For the $qqqq\bar{q}$ components in the $N^{*}(1535)$
resonance, we consider the contributions of the lowest and
next-to-lowest energy five-quark components, as in our recent
work$^{[4]}$.

\parbox[t]{200mm}
{\centerline{\psfig{figure=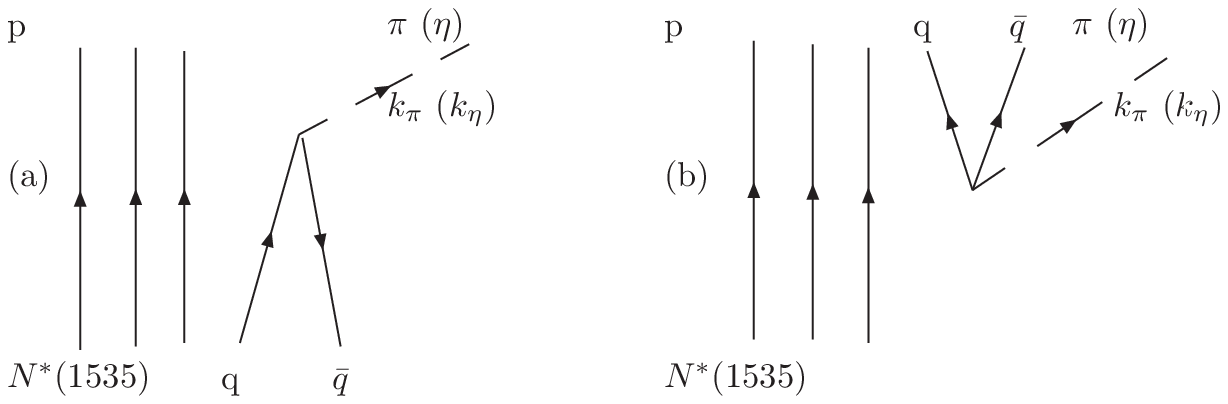}}
\centerline{\footnotesize Figure 1. \quad The $q\bar{q}\rightarrow
\pi(\eta)$ process. }} \label{non-diagonal}
\parbox[t]{200mm}
{ }

Given the presence of the $qqqq\bar{q}$ components in both the
nucleon and $N^{*}(1535)$ resonance, the strong decays of the latter
may contain the following three processes: transition between the
$qqq$ components, transition between the $qqqq\bar{q}$ components,
and the annihilation transitions between the $qqq$ and $qqqq\bar{q}$
components. The results show that the five-quark components have
significant contributions to the $N\pi$ and $N\eta$ decays of
$N^{*}(1535)$. And with these contributions, the description for the
strong decays of the $N^{*}(1535)$ resonance may be improved,
especially for the $N\eta$ decay.

The present manuscript is organized as follows: The formalism for
the strong decays of $N^{*}(1535)$ in the chiral quark model is
given in Section~\ref{sec:2}. In Section~\ref{sec:3}, we calculate
the decay width for the strong decays $N^{*}(1535)\to N\pi$ and
$N^{*}(1535)\to N\eta$ by including all the contributions of the
three- and five-quarks components. Finally, Section~\ref{sec:4}
contains a concluding discussion.

\section{Formalism for the strong decays of $N^{*}(1535)$ in the chiral quark model}
\label{sec:2}

In the chiral quark model, the coupling of the baryons to the octet
of light pseudoscalar mesons takes the form
\begin{equation}
\mathcal{L}_{Mqq}=i\frac{g^{q}_{A}}{2f_{M}}\bar{\psi}_{q}\gamma_{5}\gamma_{\mu}
\partial^{\mu}\phi_{M}X_{M}^{q}\psi_{q}\,.\label{lag1}
\end{equation}
Here $g^{q}_{A}$ denotes the axial coupling constant for the
constituent quarks, the phenomenological value for which is in the
range $0.70-1.26$$^{[10]}$. And $f_{M}$ is the decay constant of the
corresponding meson. For the $\pi$ and $\eta$ mesons, the empirical
values of the decay constants are $f_{\pi}=93$ MeV and
$f_{\eta}=112$ MeV, \vspace{5cm}

\noindent respectively. And $\psi_{q}$ and $\phi_{M}$ denote the
quark and meson field, respectively. Finally, $X_{M}^{q}$ is the
flavor operator for emission of the meson $M$ ($\pi$, $\eta$) from
the corresponding quark $q$, taking the following form:
\begin{eqnarray}
X_{\pi_{0}}^{q}&=&\lambda_{3}\nonumber\,,\\
X_{\eta_{8}}^{q}&=&\lambda_{8}\nonumber\,,\\
X_{\eta_{1}}^{q}&=&\sqrt{\frac{2}{3}}\mathcal{I}\,,\
\end{eqnarray}
where $\lambda_{i}$ are the $SU(3)$ Gell-Mann matrices, and
$\mathcal{I}$ denotes the unit operator in the flavor space. Note
that the $\eta$ meson is an admixture of the octet $\eta_{8}$ and
the singlet $\eta_{1}$ with the mixing angle $\theta_{p}$:
\begin{eqnarray}
\eta=\eta_{8}\cos{\theta_{p}}-\eta_{1}\sin{\theta_{p}}\nonumber\,,\\
\eta'=\eta_{8}\sin{\theta_{p}}+\eta_{1}\cos{\theta_{p}}\,.\
\end{eqnarray}
The value for $\theta_{p}$ is $-23^{\circ}$$^{[1]}$. Then the flavor
operator for emission of the $\eta$ meson from the corresponding
quark should be
\begin{equation}
X_{\eta}^{q}=\cos{\theta_{p}}X_{\eta_{8}}^{q}-\sin{\theta_{p}}X_{\eta_{1}}^{q}\,.\
\end{equation}

Taking the $qqqq\bar{q}$ components into account, the decays of a
baryon can be divided into three parts: (1) The transitions between
the three-quarks components, {\sl i.e.}, the process $qqq\to qqqM$,
(2) The transitions between the five-quark components, {\sl i.e.},
$qqqq\bar{q}\to qqqq\bar{q}M$, and (3) The decays through $q\bar{q}$
annihilation, {\sl i.e.}, $qqqq\bar{q}\to qqqM$ and $qqq\to
qqqq\bar{q}M$. The first process has been analyzed by many
authors$^{[11-13]}$, and it can be applied to part (2) directly. And
Figure\ref{non-diagonal} shows the last processes.

In the non-relativistic approximation, the Lagrangian~(\ref{lag1})
leads to the following operators
\begin{eqnarray}
T_{d}&=&-i\sum_{q=1}^{n_{q}}\frac{g^{q}_{A}}{2f_{M}}\exp\{-i\vec{k}_{M}\cdot\vec{r}_{q}\}
\vec{\sigma}_{q}\cdot(\vec{k}_{M}\nonumber\\
&&-\omega_{M}\frac{\vec{p}_{q}}{m_{q}})X^{q}_{M}\nonumber\,,\\
T_{nd}&=&i\sum_{q=1}^{4}\frac{g^{q}_{A}m_{q}}{f_{M}}\exp\{-i\vec{k}_{M}\cdot(\vec{r}_{q}+\vec{r}_{5})/2\}\nonumber\\
&&\delta^{3}(\vec{r}_{q}-\vec{r}_{5})X^{q}_{M}\,.\ \label{operator}
\end{eqnarray}
Here $\hat{T}_{d}$ denotes the diagonal transition operator, {\sl
i.e.}, the operator for parts (1) and (2), the sum runs over all the
quarks in the corresponding components of the nucleon and
$N^{*}(1535)$, which means that $n_{q}=3$ and $n_{q}=5$ for the
transitions between the three-quark components and five-quark
components in the nucleon and $N^{*}(1535)$, respectively. And
$\hat{T}_{nd}$ denotes the non-diagonal transition operator, {\sl
i.e.}, the operator for part (3), the sum runs over the four-quarks
of the five-quark components in the nucleon or $N^{*}(1535)$. And
$\vec{\sigma}_{q}$ denotes the spin operators of the quark which
emits a meson, $\vec{r}_{q}$ and $m_{q}$ the coordinate and
constituent mass of the corresponding quark, and $\vec{r}_{5}$ the
coordinate of the anti-quark. Finally $\vec{k}_{M}$ is the momentum
of the meson.

Once the transition amplitude of the $N\pi$ and $N\eta$ decays are
obtained, one can calculate the decay width directly by using the
following formula
\begin{eqnarray}
d\Gamma_{N^{*}(1535)\rightarrow
N\pi}&=&\frac{3}{16\pi^{2}}\frac{E^{'}+m_{N}}{m^{*}}|\vec{k}_{\pi}|
|T^{(\pi)}|^{2}d\Omega\,,\nonumber\\
d\Gamma_{N^{*}(1535)\rightarrow
N\eta}&=&\frac{1}{16\pi^{2}}\frac{E^{'}+m_{N}}{m^{*}}|\vec{k}_{\eta}|
|T^{(\eta)}|^{2}d\Omega\,,\nonumber\\
&&  \label{width}
\end{eqnarray}
here $E^{'}$ and $m_{N}$ are the energy and mass of the nucleon,
respectively, and $m^{*}$ the mass of the resonance $N^{*}(1535)$.
And $\vec{k}_{\pi}$ and $\vec{k}_{\eta}$ denote the three-momentum
of the $\pi$ and $\eta$ mesons, respectively. And in the present
case,
\begin{eqnarray}
E^{'}&=&\frac{m^{*2}-m_{M}^{2}+m_{N}^{2}}{2m^{*}}\,,\nonumber\\
|\vec{k}_{M}|&=&\{[m^{*2}-(m_{N}+m_{M})^{2}][m^{*2}-(m_{N}\nonumber\\
&&-m_{M})^{2}]\}^{1/2}/2m^{*}\,.\
\end{eqnarray}
For the $N\pi$ decay, $k_{\pi}=467$ MeV and $E^{'}=1048$ MeV; and
for the $N\eta$ decay, $k_{\eta}=188$ MeV and $E^{'}=957$ MeV.

\section{The numerical results}
\label{sec:3}

\subsection{Contributions of $qqq$ components}
\label{subsec:1}

At the first step we consider the diagonal transitions of the
three-quark components $qqq\rightarrow qqqM$. The wave functions of
the corresponding components are taken to be the conventional
ones$^{[6]}$. With these wave functions given in Ref.$^{[6]}$ and
the operator~(\ref{operator}), a straightforward calculation leads
to the following matrix elements of the operator $\hat{T}_{d}$
between the three-quark components in the proton and $N^{*}(1535)$:
\begin{eqnarray}
\langle\hat{T}_{d}^{(\pi_{0})}\rangle_{(3q)}&=&-iA_{N3}A_{N^{*}3}\frac{g^{q}_{A}}
{2f_{\pi}}\frac{2\sqrt{2}}{9}\omega_{3}[\frac{k_{\pi_{0}}^{2}}{\omega_{3}^{2}}
(1-\frac{\omega_{\pi_{0}}}{3m})\nonumber\\
&&-\frac{3\omega_{\pi_{0}}}{m}]\exp\{-\frac{k_{\pi_{0}}
^{2}}{6\omega_{3}^{2}}\}\nonumber\,,\\
\langle\hat{T}_{d}^{(\eta)}\rangle_{(3q)}&=&-iA_{N3}A_{N^{*}3}\frac{g^{q}_{A}}{2f_\eta}
\frac{\sqrt{6}}{9}\omega_{3}C_{\eta}[\frac{k_{\eta}^{2}}{\omega_{3}^{2}}
(1-\frac{\omega_{\eta}}{3m})\nonumber\\
&&-\frac{3\omega_{\eta}}{m}]\exp\{-\frac{k_{\eta}
^{2}}{6\omega_{3}^{2}}\}\,.\
\end{eqnarray}
Here $m$ is the constituent mass of the light quark, and the factor
$C_{\eta}=\cos{\theta_{p}}-\sqrt{2}\sin{\theta_{p}}$, which comes
from the mixing of the $\eta_{1}$ and $\eta_{8}$ mesons. Then we can
get the decay widths for the $N\pi$ and $N\eta$ decays of
$N^{*}(1535)$ employing q.(\ref{width}). The numerical results are
shown in Table~\ref{table1}, with the axial coupling constant
$g^{q}_{A}=0.70$, $g^{q}_{A}=0.75$ and $g^{q}_{A}=0.80$,
respectively. At the present step, we have not considered the
five-quark components, so the amplitudes of the three-quark
components in the nucleon and $N^{*}(1535)$ should be
$A_{N3}=A_{N^{*}3}=1$. The results for these decay widths are
proportional to $(g^{q}_{A})^2$, as shown in Table~\ref{table1}. We
will use the central value $g^{q}_{A}=0.75$ for our further
calculations. The three columns in Table~\ref{table1} are obtained
by setting the oscillator parameter to be $\omega_{3}=340$ MeV,
which is employed for the electromagnetic transition
$\gamma^{*}N\rightarrow N^{*}(1535)$ in Ref.$^{[4]}$, and the
constituent mass of the light quark has been taken to be $m=340$
MeV. The experimental data for the decay width are extracted from
Ref.$^{[1]}$, here we have taken the average value for the
Breit-Wigner width.

\begin{center}
{\footnotesize\bf Table 1\quad The decay widths for $N^{*}(1535)\to
N\pi$, $N\eta$ without including $qqqq\bar q$ components. The
results (A-C) correspond to the axial coupling constant $g^{q}_{A}$
to be $0.70$, $0.75$ and $0.80$, respectively, with the oscillator
parameter $\omega_{3}=340$ MeV.}\vspace{1.3mm}\\
\footnotesize \doublerulesep 0.4pt \tabcolsep 6pt
\begin{tabular}{ccccc}
\hline \hline &$A $& $B $& $C $& Data \\
$\Gamma_{N^{*}\rightarrow N\pi}$ (MeV)    &   136.7  &    156.9    &     178.6     &  52.5-82.5   \\

$\Gamma_{N^{*}\rightarrow N\eta}$ (MeV)   &   44.2   &     50.7    &      57.7     & 67.5-90.0    \\

$R=\frac{\Gamma_{N^{*}\rightarrow N\eta}}{\Gamma_{N^{*}\rightarrow
N\pi}}$                                       &  0.32    &   0.32     &    0.32     & 0.82-1.71   \\

\hline \hline\hline
\end{tabular}
\label{table1}
\end{center}\vspace{1mm}

As shown in Table~\ref{table1}, all of the three columns cannot fit
the data well. For instance, in column A, the decay width for the
$N\pi$ decay of $N^{*}(1535)$ is much larger than the upper bound of
the experimental data, while that for the $N\eta$ decay is a bit
smaller than the lower bound of the data. Consequently, the value
for the ratio $R=\frac{\Gamma_{N^{*}(1535)\rightarrow
N\eta}}{\Gamma_{N^{*}(1535)\rightarrow N\pi}}$ is much smaller than
the lower bound of the data in Ref.$^{[1]}$. It indicates that we
should consider the contributions of the five-quark components in
the nucleon and $N^{*}(1535)$.

\subsection{Contributions of the five-quark components with the lowest energy} \label{subsec:2}

Then we should consider the contributions of the $qqqq\bar{q}$
components. As mentioned in Section~\ref{subsec:1}, it can be
divided into two parts: the diagonal and non-diagonal transitions.
For the former one, the operator is $\hat{T}_{d}$ in
Eq.(\ref{operator}) with $n_{q}=5$. First, we consider the
$qqqq\bar{q}$ components in the nucleon and $N^{*}(1535)$ with the
lowest energy in this section.

The orbital-flavor-spin configuration for the four-quark subsystem
of the lowest energy five-quark component in $N^{*}(1535)$ is
$[4]_{X}[31]_{FS}[211]_{F}[22]_{S}$, and that for the four-quark
subsystem of the lowest energy five-quark components in the nucleon
is $[31]_{X}[4]_{FS}[22]_{F}[22]_{S}$. Note that the total spin of
the four-quark subsystem is then $S=0$; consequently, the matrix
element of $\hat{T}_{d}$ between these four quarks vanishes. On the
other side, the matrix element of $\hat{T}_{d}$ between the
anti-quarks in the nucleon and $N^{*}(1535)$ should also vanish for
the orthogonality of the orbital states $[4]_{X}$ and $[31]_{X}$.
Then the diagonal transition between the lowest energy $qqqq\bar{q}$
components in $N^{*}(1535)$ and nucleon does not contribute to the
$N\pi$ and $N\eta$ decays of the $N^{*}(1535)$ resonance.

On the other hand, calculations of the matrix elements for the
operator $\hat{T}_{nd}$ between the wave functions of the proton and
$N^{*}(1535)$ lead to the following expressions:
\begin{eqnarray}
\langle\hat{T}_{nd}^{(\pi_{0})}\rangle_{(3q5q)}&=&0\nonumber\,,\\
\langle\hat{T}_{nd}^{(\eta)}\rangle_{(3q5q)}&=&iA_{N3}A_{N^{*}s\bar{s}}
\frac{g^{q}_{A}m_{s}}{f_{\eta}}\frac{\sqrt{6}}{6}C_{35}C'_{\eta}\nonumber\\
&&\exp\{-\frac{3k_{\eta}^{2}}{20\omega_{5}^{2}}\}\,.\ \label{lowest}
\end{eqnarray}
Here $C_{35}$ denotes the orbital overlap factor:
\begin{equation}
\langle\varphi_{00}(\vec{\kappa}_{1})\varphi_{00}(\vec{\kappa}_{2})
|\varphi_{00}(\vec{\kappa}_{1})\varphi_{00}(\vec{\kappa}_{2})\rangle\
=(\frac{2\omega_{3}\omega_{5}}{\omega_{3}^{2}+\omega_{5}^{2}})^{3}
\, .
\end{equation}
And the factor
$C'_{\eta}=\sqrt{2}\cos{\theta_{p}}+\sin{\theta_{p}}$.

\begin{center}
{\footnotesize\bf Table 2\quad The decay widths for $N^{*}(1535)\to
N\pi$, $N\eta$ by including the lowest $qqqq\bar q$ components with
$g^{q}_{A}=0.75$. The results (A-E) correspond to the portion of
$qqqq\bar q$ components in
$N^{*}(1535)$ to be 25\%, 35\%, 45\%, 55\% and 65\%, respectively.}\vspace{1.3mm}\\
\footnotesize \doublerulesep 0.4pt \tabcolsep 3pt
\begin{tabular}{ccccccc}
\hline \hline &$A $& $B $& $C $& $D $&
E& Data \\
$\Gamma_{N^{*}\rightarrow N\pi}$ (MeV)    &  152.0   &   131.7     &    111.4    &    91.2  & 70.9& 52.5-82.5   \\

$\Gamma_{N^{*}\rightarrow N\eta}$ (MeV)   &  60.9    &   57.6      &   53.3      &  48.3    &42.5& 67.5-90.0    \\

$R=\frac{\Gamma_{N^{*}\rightarrow N\eta}}{\Gamma_{N^{*}\rightarrow
N\pi}}$ &  0.40    &    0.44 &
0.48  &  0.53      & 0.60    &  0.82-1.71   \\
\hline\hline
\end{tabular}
\label{table2}
\end{center}\vspace{1mm}

As we can see from Eq.(\ref{lowest}), the lowest energy five-quark
components in the nucleon and $N^{*}(1535)$ do not contribute to the
$N\pi$ decay of $N^{*}(1535)$, while they contribute to the $N\eta$
decay. The numerical results are shown in Table~\ref{table2}. The
proportion of the five-quark component in $N^{*}(1535)$ is taken to
be $25-65\%$, and in the nucleon $20\%$. And the oscillator
parameters are taken to be $\omega_{3}=340$ MeV and $\omega_{5}=600$
MeV, with which values we can describe the electromagnetic
transitions $\gamma^{*}N\rightarrow N^{*}(1535)$ best$^{[4]}$. Since
we have considered the five-quark components in the nucleon and
$N^{*}(1535)$, we should set the constituent quark masses to be
smaller than the one employed in Section~\ref{subsec:1}. Here we
take $m=290$ MeV for the light quarks and $m_{s}=430$ MeV for the
strange quark.

As shown in Table~\ref{table2}, the numerical result for the $N\eta$
decay of $N^{*}(1535)$ is getting better when the contributions of
the five-quark components are taking into account, and the ratio $R$
of the $N\eta$ and $N\pi$ decay rates increases. Note that here we
have only considered the contributions from the lowest energy
five-quarks component in $N^{*}(1535)$, which contains only the
strange $s\bar{s}$ quark pair and hence does not contribute to the
$N\pi$ decay. The numerical results are still not good enough.
Therefore, the contributions from other flavor-spin configurations,
which allow for the five-quark components with light $q\bar{q}$,
need to be considered. And these configurations may contribute to
both $N\pi$ and $N\eta$ decays of $N^{*}(1535)$. As in Ref$^{[4]}$,
we consider the flavor-spin configuration
$[31]_{FS}[22]_{F}[31]_{S}$ for the four-quark subsystem of the
five-quark components, which has the next-to-lowest energy.

\subsection{Contributions from other configurations}
\label{subsec:3}

As mentioned in Section~\ref{subsec:2}, here we consider the
flavor-spin configuration $[31]_{FS}[22]_{F}[31]_{S}$ for the
four-quark subsystem of the five-quark components in $N^{*}(1535)$,
which has the next-to-lowest energy allowing for the $qqqq\bar{q}$
components with both light and strange quark pairs. The explicit
wave functions for this configuration have been given in
Ref.$^{[4]}$. And straightforward calculations of the matrix
elements of the operator $\hat{T}_{d}$ in Eq.(\ref{operator}) lead
to the following transition amplitudes
\begin{eqnarray}
\langle\hat{T}_{d}^{(\pi_{0})}\rangle_{(5q5q)}\!&\!\!=\!\!&\!
iA_{Ns\bar{s}}A'_{N^{*}s\bar{s}}\frac{g^{q}_{A}}{f_{\pi}}\frac{\sqrt{2}}{24}
\frac{\omega_{\pi}\omega_{5}}{m}
\exp\{-\frac{k_{\pi}^{2}}{5\omega_{5}^{2}}\}\,,\nonumber\\
\langle\hat{T}_{d}^{(\eta)}\rangle_{(5q5q)}\!&\!\!=\!\!&\!
iA_{Nd\bar{d}}A_{N^{*}d\bar{d}}\frac{g^{q}_{A}}{f_{\eta}}\frac{\sqrt{6}}{18}
c_{\eta}\frac{\omega_{\eta}\omega_{5}}{m}
\exp\{\nonumber\\
-\frac{k_{\pi}^{2}}{5\omega_{5}^{2}}\}\!&\!\!+\!\!&\!iA_{Ns\bar{s}}A'_{N^{*}s\bar{s}}\frac{g^{q}_{A}}{2f_{\eta}}
(\frac{\sqrt{6}}{12}c_{\eta}\frac{\omega_{\eta}}{m}\nonumber\\
&&-\frac{\sqrt{3}}{18}
c'_{\eta}\frac{\omega_{\eta}}{m_{s}})\omega_{5}
\exp\{-\frac{k_{\pi}^{2}}{5\omega_{5}^{2}}\}\,. \label{d5}
\end{eqnarray}
On the other hand, calculations of the matrix elements of the
operator $\hat{T}_{nd}$ lead to the following non-diagonal
transition amplitudes
\begin{eqnarray}
\langle\hat{T}_{nd}^{(\pi_{0})}\rangle_{(3q5q)}\!&\!\!=\!\!&\!-iA_{N3}A_{N^{*}d\bar{d}}\frac{2g^{q}_{A}m}
{f_{\pi}}C_{35}
\exp\{-\frac{3k_{\eta}^{2}}{20\omega_{5}^{2}}\}\,,\nonumber\\
&&
\end{eqnarray}

\begin{eqnarray}
\langle\hat{T}_{nd}^{(\eta)}\rangle_{(3q5q)}&=&iA_{N3}A_{N^{*}d\bar{d}}\frac{g^{q}_{A}m}
{f_{\eta}}\frac{2}{\sqrt{3}}C_{35}C_{\eta}
\exp\{\nonumber\\
&&-\frac{3k_{\eta}^{2}}{20\omega_{5}^{2}}\}-iA_{N3}A'_{N^{*}s\bar{s}}\frac{g^{q}_{A}m_{s}}
{f_{\eta}}\frac{2}{\sqrt{3}}\nonumber\\
&&C_{35}C'_{\eta} \exp\{-\frac{3k_{\eta}^{2}}{20\omega_{5}^{2}}\}.
\label{nd5}
\end{eqnarray}

Note as in Ref.$^{[4]}$, we have not included explicitly the
contributions of the next-to-lowest energy five-quarks components
containing $s\bar s$ pair, because these contributions can be
replaced by certain probabilities of the lowest energy five-quarks
component in $N^{*}(1535)$.

Including contributions from the five-quark components of the
next-to-lowest energy allowing for the light $q\bar q$ pairs, the
numerical results are shown in Table~\ref{table3}, with the
oscillator parameters $\omega_{3}=340$ MeV and $\omega_{5}=600$ MeV,
and the axial coupling constant $g^{q}_{A}=0.75$. Here we take the
probability $P_{5q}$ for the five-quarks components in $N^{*}(1535)$
to be $25\%$ to $65\%$, the proportion of the lowest energy
$qqqq\bar{q}$ component to be $0.9P_{5q}$, and that of the
next-to-next-to-lowest energy one to be $0.1P_{5q}$. As shown in
Table~\ref{table3}, the numerical results with $P_{5q}> 40\%$ fall
well in the range of the data with uncertainties.

\begin{center}
{\footnotesize\bf Table 3\quad The decay widths for $N^{*}(1535)\to
N\pi$, $N\eta$ by including the lowest and next-to-lowest $qqqq\bar
q$ components with $g^{q}_{A}=0.75$. The results (A-E) correspond to
the portion of $qqqq\bar q$ components in
$N^{*}(1535)$ to be 25\%, 35\%, 45\%, 55\% and 65\%, respectively.}\vspace{1.3mm}\\
\footnotesize \doublerulesep 0.4pt \tabcolsep 3pt
\begin{tabular}{ccccccc}
\hline \hline &$A $& $B $& $C $& $D $&
E& Data \\
$\Gamma_{N^{*}\rightarrow N\pi}$ (MeV)    &  111.9   &    88.5     &    67.4  &    48.4    &  31.4  & 52.5-82.5   \\

$\Gamma_{N^{*}\rightarrow N\eta}$ (MeV)   &  59.9    &   56.4      &   52.1   &  47.0       &  41.2    & 67.5-90.0    \\

$R=\frac{\Gamma_{N^{*}\rightarrow N\eta}}{\Gamma_{N^{*}\rightarrow
N\pi}}$ &  0.54    &    0.64 &
0.77    &  0.97    &  1.31        & 0.82-1.71   \\

\hline \hline\hline
\end{tabular}
\label{table3}
\end{center}\vspace{1mm}

The portion of the five-quarks components in $N^{*}(1535)$ were
predicted to be $25-65\%$ in Ref.$^{[4]}$. While the results in
Table~\ref{table3} correspond to the full width of 150 MeV for
$N^*(1535)$, the width is in fact not well determined in the range
of $100-200$ MeV $^{[1]}$. With such uncertainty one can get good
descriptions for the strong decays of $N^{*}(1535)$ with the
proportion of the $qqqq\bar{q}$ components in the range $45-65\%$.
Consequently, with the portion of the five-quarks components in
$N^{*}(1535)$ in the range $45-65\%$, we can describe both of the
electromagnetic and strong decays of $N^{*}(1535)$ well in the same
model.

\section{Conclusion}
\label{sec:4}

The role of the low-lying $qqqq\bar{q}$ components in the strong
decays of the $N^{*}(1535)$ is investigated. The orbital-flavor-spin
configuration for the four-quark subsystem in the $qqqq\bar{q}$
components in the nucleon is assumed to be
$[31]_{X}[4]_{FS}[22]_{F}[22]_{S}$, which is expected to have the
lowest energy for the five-quark component, and therefore most
likely to form appreciable components in the proton. In the case of
the $N^{*}(1535)$ resonance, the lowest energy configuration is
$[4]_{X}[31]_{FS}[211]_{F}[22]_{S}$, which only allows five-quark
component with $s\bar{s}$ component. In addition, we have also
considered the next-to-lowest energy configuration
$[4]_{X}[31]_{FS}[22]_{F}[31]_{S}$, allowing for the five-quark
components with both light and strange $q\bar{q}$ pairs.

The results show that the lowest energy $qqqq\bar{q}$ configuration
does not contribute to the $N\pi$ decay of $N^{*}(1535)$, while it
has nonzero contributions to the $N\eta$ decay. And the
contributions come mainly from the annihilation transition
$qqqs\bar{s}\to qqq$, which depends on the portion of $qqqq\bar{q}$
admixture in $N^{*}(1535)$ and the oscillator parameters
$\omega_{3}$ and $\omega_{5}$. Here we take the oscillator
parameters to be the same as the ones used in Ref.$^{[4]}$, which
gives a good description for the electromagnetic transition
$\gamma^{*}N\to N^{*}(1535)$.

The five-quark components with next-to-lowest energy five-quark
components have nonzero contributions to both $N\pi$ and $N\eta$
decays of $N^{*}(1535)$, through both diagonal and non-diagonal
transitions. Considering the contributions of the lowest and
next-to-lowest energy five-quark components, we can give a
description of the $N\pi$ and $N\eta$ decays consistent with data.
Consequently, both electromagnetic and strong decays of
$N^{*}(1535)$ can be described well by the same model, and with the
same parameters.

All above suggest that there are large five-quark components in
$N^{*}(1535)$, which may contribute significantly to the strong
decays of this resonance. In addition, we conclude that it should be
instructive to extend this phenomenological analysis to the case of
all baryon resonances. Accumulation of data on the baryon resonances
from BEPC2, CSR and other facilities will be extremely helpful for
our understanding the internal structures of these baryons.

\bigskip

We thank Prof. En-Guang Zhao for encouragements on our research, and
Prof. D. O. Riska and Dr H. Q. Zhou for helpful discussions.


\end{multicols}

\begin{multicols}{2}
\normalsize \vskip0.16in\parskip=0mm \baselineskip 15pt
\renewcommand{\baselinestretch}{1.12}
\footnotesize
\parindent=4mm

\bahao\REF{1\ } Particle Data Group, Amsler C et al. Phys. Lett. B
667: 1 (2008)

\bahao\REF{2\ } Xie J J, Zou B S, Chiang H C. Phys. Rev. C 77:
015206 (2008)

\bahao\REF{3\ } Liu B C, Zou B S. Phys. Rev. Lett 96: 042002 (2006);
Phys. Rev. Lett. 98: 039102 (2007)

\bahao\REF{4\ } An C S, Zou B S. Eur. Phys. J. A 39: 195-204 (2009)

\bahao\REF{5\ } Glozman L Y, Riska D O. Phys. Repts 268: 263-303
(1996)

\bahao\REF{6\ } Riska D O, Brown G E. Nucl. Phys. A 679: 577-596
(2001)

\bahao\REF{7\ } Li Q B, Riska D O. Phys. Rev. C 73: 035201 (2006)

\bahao\REF{8\ } Li Q B, Riska D O. Phys. Rev. C 74: 015202 (2006)

\bahao\REF{9\ } An C S, Riska D O, Zou B S. Phys. Rev. C 73: 035207
(2006)

\bahao\REF{10\ } Goity J L, Roberts W. Phys. Rev. D 60: 034001
(1999)

\bahao\REF{11\ } Koniuk R, Isgur N. Phys. Rev D 21: 1868 (1980)

\bahao\REF{12\ } Capstick S, Roberts W. Phys. Rev. D 47: 1994 (1993)

\bahao\REF{13\ } Capstick S, Roberts W. Pro. Par. Nucl. Phys. 45:
S241-S331 (2000), and references therein

\end{multicols}
\tlj